\documentclass[twocolumn,twoside,preprintnumbers,amsmath,amssymb,showkeys]{revtex4}

\usepackage{amsmath,amssymb}
\usepackage{graphicx}
\usepackage{epsf}
\usepackage{epsfig}
\usepackage[figuresright]{rotating}

\usepackage{fancyhdr}
\usepackage{pslatex}

\newcommand{\be}{\begin{eqnarray}}

\newcommand{\ee}{\end{eqnarray}}

\begin{document}

\title{Summary Talk for ISMD 06\\}
\author{Larry McLerran }
\affiliation{ Physics Department and Riken Brookhaven Center,
PO Box 5000, Brookhaven National 
Laboratory,  
 Upton, NY 11973 USA}

\begin{abstract}
I give an overview of the presentations at the International Symposium
on Multiparticle Dynamics. 2006
\end{abstract}
\maketitle
\section{The Questions}

I think it is appropriate to ask what are some of the "Big Questions" in particle and nuclear physics,
and how does this meeting address them.  The choice of questions
reflects my view of the field, and it is certainly not universally held.
\begin{itemize}
\item{\bf What are the fundamental interactions?}

We understand electromagnetic interactions, and weak interactions up to
a scale of around $100~GeV$.  Classical gravity is understood, and
quantum gravity is the subject of much theoretical speculation with little experimental
input.  

QCD is the central subject of this meeting.  The basic interactions of QCD are understood,
and one has a wide variety of experimental tests of QCD at high energy and large momentum
transfer.  This is the short distance limit of the theory.

\item{\bf What is the structure of matter?}

Strongly interacting matter can take many forms.  Strong interactions bind the quarks and gluons
into hadrons, and make nuclei from nucleons.  Part  of the subject of this
meeting is how matter is formed from the collisions of particles at very high energy.
This involves the formation of quarks and gluons from energy in the initial collision process, and
ultimately the hadronization of quarks and gluons.

\item{\bf What are the different forms this matter can take?}

One of the remarkable features of the high energy limit of QCD is that it appears to
be described by new forms of matter.  The part of the wavefunction which controls
the high energy limit of QCD is composed of gluons in a very high energy density, highly coherent state,
the Color Glass Condensate.  This mattes is liberated upon collision, and has properties
like a plasma except with high density color electric and magnetic fields, the so called Glasma.
In nuclear collisions, and also possibly in pp collisions at the highest energies, this matter thermalizes
and forms a Quark Gluon Plasma.\cite{cgcreview}-\cite{zakopane}  

These new forms of matter, and perhaps other forms of  matter not yet thought of,  are probed by studying the multiparticle dynamics of 
high energy collisions.  Insofar as this matter has universal properties, the study of this matter is of fundamental interest.

In order to understand the matter formed in these collisions, we need to develop fully the space-time
description of high energy collisions. This understanding is good in some cases, such
as the formation of jets, or the relatively late stage evolution of a Quark Gluon Plasma.  It
is the subject of much more speculation when describing hadronization, or the early time
development of a Glasma.
\end{itemize}

Before proceeding, I should say that it is impossible to be fully comprehensive in a summary talk.
It is even harder in the written version of the talk, as there is even a tighter space limitation.
So please forgive me if I have not discussed, nor fully elaborated, on the subject of your presentation.
Also, please be tolerant of my lack of deep understanding of many of the topics presented here.
I also have taken the liberty of referring to the original literature collectively through
some very nice review papers.  Please find references to the original literature in these 
reviews.\cite{cgcreview}-\cite{whitepaper}
The contributions to the conference which I discuss below are
of course part of these conference proceedings.\cite{ismd}  Also, please look for references to some of the original
literature described in these talks in the corresponding written contributions.

\section{Beyond the Standard Model}

There were two talks at this meeting about the physics beyond the standard 
model.  the first by Bill Gary concerned CP violation in B decays, $ B^0 \rightarrow \overline K^{*0} K^0$.
One of the reasons for this study is to test the unitarity of the CKM model of CP violation.
If a lack of unitarity was found, then this would imply that the Standard Model must be extended.
There are a variety of inputs needed to draw such a conclusion, including high precision
lattice computations of weak matrix elements.

The other talk was by Horst Stoecker, who has been using extra dimensions and TeV scale gravity
to explore the possible new particles which might be produced at the LHC.  This work is
very speculative, since extra dimensions might not exist, and if they did, the size of such
an extra dimension could be anywhere from the Compton wavelength associated with the TeV energy scale up to the Planck length.  There are also a virtual continuum of mass ranges and properties
of the particles predicted by extra dimensions.  The test of all of this
will be the LHC.  If there is something there associated with TeV mass black holes, theoretical
physics will be changed in a deep and fundamental way.  If there is nothing there, few of us would lose sleep.  Such is the nature of speculation.  

\section{QCD Works at Short Distances}

\begin{figure}[ht]
    \begin{center}
        \includegraphics[width=0.45\textwidth]{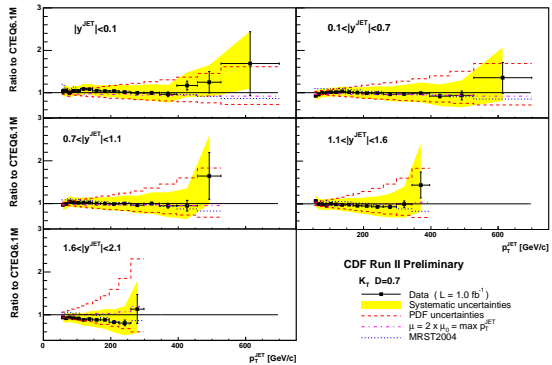}
        \caption{A Comparison of pQCD computations with CDF jet production data. }
\label{mesropian}
    \end{center}
\end{figure}

 QCD describes strong interaction processes well at short distances.  One of course has to choose
 infrared safe observables, and for many jet computations one typically uses Monte-Carlo fragmentation
 models.  In the presentation by Maramidas, the cross section predicted by NLO pQCD was shown to
 describe well the D production data in deep inelastic scattering measured by Zeus. In the presentation
 by Mesropian, CDF jet production was compared to pQCD computations.  This is shown in Fig.
 \ref{mesropian}

There were presentations showing the agreement of NLO computations with photon-gluon fusion
and photon-quark scattering to produce pairs of jets
in deep inelastic scattering, by Efremenko, as seen in the H1 experiment..  Messinia from CDF presented data on associated jet production in the production of W bosons.  In both cases the agreement with pQCD is quite good.  Soares presented data on mutlti-jet production at Zeus.  In this case, the agreement is not good as the other pQCD comparisons, but this is presumably because
the theory computations have not been done with corresponding accuracy.

The remarkable agreement between pQCD and short distance processes in QCD forces us
to ask:  

\vspace{0.1in}
\noindent{\bf Where are the frontiers of our knowledge?}
\vspace{0.1in}

One of the issues here is that the distribution and fragmentation functions in QCD are largely phenomenological.  One can use DGLAP equations to describe their evolution, for high
$Q^2$ and not too small x.  Perhaps the BFKL evolution works at small $x$ and not too
large $Q^2$.  How these distribution functions originate either from boundary conditions
in solving the evolution equations, or as universal fixed points of the evolution equations is 
not fully understood.
This means that as a matter of first principle in QCD, we neither have full understanding of the 
the origin of the distribution quarks and gluons inside a hadron wavefunction nor how hadrons are produced
from these quarks and gluons.  

There are also a wide variety of machines, some operating, some almost operating,
and some proposed which can help us to understand these issues.  Hera has provided hints
about the nature of matter which controls the physics at the highest energies.  This comes 
from both deep inelastic scattering and diffractions at small values of x.  RHIC
has produced a strongly interacting Quark Gluon Plasma, and we are beginning to learn
some of its properties.  It has also provided hints about the nature of small x matter produced
at Hera, and about how this matter is converted from the wavefunction of a nucleus into
matter which evolves and ultimately becomes the Quark Gluon Plasma.  If the theoretical
speculations concerning these results from RHIC and Hera are more or less correct,
we have (in my opinion) the beginnings of a first principles understanding of the high energy limit of QCD,
and the remarkable conclusion that it is due to the universal properties of matter made 
in these collisions. 
Soon, we will have the LHC with unprecedented range in 
$x$ and $Q^2$, with potential for both new discovery, and perhaps turning some of the hints seen
at Hera and RHIC into substantial scientific discovery.  An electron-ion collider dedicated to
QCD studies, may provide detailed quantitative tests of hypothesis about the nature of such
matter.

\section{Deep Inelastic Scattering and Diffraction}
 
\begin{figure}[ht]
    \begin{center}
        \includegraphics[width=0.45\textwidth]{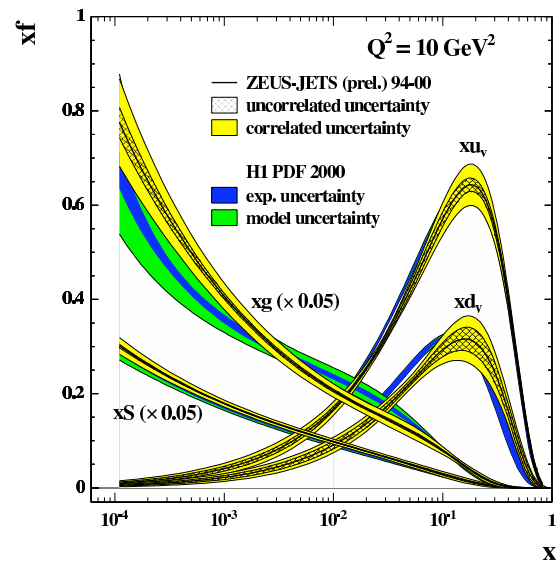}
        \caption{The gluon distribution function measured at Hera. }
\label{glue}
    \end{center}
\end{figure}

 Glazov presented the latest results from Hera on the distributiuon of quarks and gluons seen at Hera.
 In Fig. \ref{glue}, the distribution of gluons extracted from the measurements of quarks is shown.  
The rise of the gluon density at small x has led to the idea that the gluon density gets quite
large.  Ultimately, at any fixed $Q^2$, general arguments require the gluon density
to stop growing so rapidly with decreasing $x$.  This phenomenon is called saturation.
It is a a consequence of the repulsive interactions of a high density state of gluons.
Much theoretical speculation as to the nature of this saturated matter has arisen,
and the most popular of these ideas is that the high density gluons form a Color Glass Condensate.
This is a highly coherent distribution of gluons with properties similar to that of Bose condensates and
spin glasses.\cite{cgcreview}  

\begin{figure}[ht]
    \begin{center}
        \includegraphics[width=0.45\textwidth]{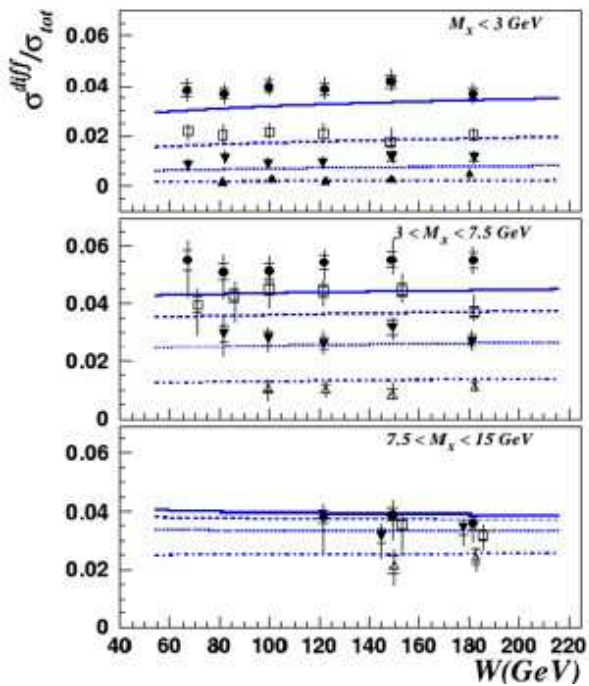}
        \caption{The ratio of deep inelastic to diffractive cross sections measured at H1}
\label{difratio}
    \end{center}
\end{figure}

Janssen presented data on the cross section for diffractive deep inelastic scattering.  Diffractive scattering is like deep inelastic scattering, except that the final state has no particles with 
$x$ values roughly between that of the photon and that of the target.  One produces a few
particles with $x$ close to that of the photon, and then there is a gap with no particles
at intermediate $x$ ranges.  One of the predictions of saturation models is that
the ratio of deep inelastic diffraction to deep inelastic scattering cross sections
is roughly constant.  Data on these cross section ratios is shown in Fig. \ref{difratio}  Polini presented data on diffractive production of vector mesons.

Kugeratiski presented a theoretical analysis of the diffractive data within a saturation
model based on ideas related to the Color Glass Condensate.\cite{zakopane}  He argued that diffractive
scattering on nuclei provide sensitive tests of these ideas.  The physical reason for this is
that saturation predicts a black disk scattering limit at the $Q^2$values for which
there is saturation.  Diffractive scattering is strongest when scattering from black disks.
Nuclei, since they have higher gluon densities, allow one to probe at higher values of
$Q^2$ where theoretical computations are better under control. 

Ducatti and Machado also presented a dipole model of deep inelastic scattering.  The cross
secion for deep inelastic scattering at small x has a scaling property
\be
\sigma_{\gamma^* p} = F(Q^2/Q_{sat}^2(x)),
\ee
 and does not have a separate
dependence on $x$.  All of the dependence on $x$ comes through the saturation momentum,
and its dependence on $x$.  This dependence may be inferred phenomenolgically,
or by theoretical computation.  The scaling property was established by Golec-Biernat, Kweicinski
and Stasto and well describes deep inelastic data for $x \le 10^{-2}$.\cite{zakopane}  Machado and Ducatti
established that gemetirc scaling works also for scattering from nuclei.

\section{Exotic Resonances}

The exotic resonances seen in BABAR at SLAC and in CLEO, Belle and Focus were the
subject of Nielsen' s talk.  She argued that these states might be interpreted as charm molecules.
Her analysis requires a reinterpretation of some low lying hadronic states as molecular
states.  This aspect of QCD, molecular bounds states or states containing glue is fascinating
as it probes the multiparticle dynamics of QCD.  It is very difficult, nevertheless since
there is always mixing between quark-antiquark states and gluons, and ithis is difficult to disentangle.

\section{Matter in the Earliest Stages of Hadronic Collisios}

The matter in the initial wavefunctions of colliding hadrons is very coherent.  That is the nature
of a bound state wavefunction.  This matter must somehow become decoherent in the scattering
process and form distributions of quarks and gluons.  In the Color Glass Condensate description,
there is a collision of two sheets of colored glass, which then melt into gluons.  During
this melting, there are highly coherent color electric and color magnetic fields.  These fields
carry a topological charge density.  
This matter thermalizes in the collisions of large nuclei, and probably also in the collisions of protons at the highest energies.  The matter at intermediate times between the initial collision and
the ultimate formation of a Quark Gluon Plasma is called the Glasma.\cite{cgcreview}-\cite{zakopane}

Of course there are alternative frameworks to that of the Color Glass Condensate, and they share
the common goals and many common features of the Color Glass Condensate description.   
Gustafson gave presentations where he used the Lund string model to attempt a generic understanding
of the formation of matter.  He argued that Lund kinematic diagrams provide an understanding
of anomalous dimensions.  He also addresses one of the unresolved problems of QCD: 
Pomeron Loops or Ploops.   Pomerons can be thought of as collective excitations of the Color Glass, or
more generally the matter in the initial state hadronic wavefunction.  In collisions, one
excites these modes.  Such modes should have quantum fluctuations, and in diagrammatic
language, these are loops.  The Gribov Reggeon Calculus was one attempt to make a theory
of Pomeron loops.  In either the Color Glass or the Lund String model, one should ultimately
have a complete theory of such Ploops.

There are haunting similarities between the Lund String Model and the Color Glass Condensate
descriptions.  Perhaps in future years, these descriptions will somehow merge.

Strikland presented arguments that during the Glasma phase, the initially approximately
boost invariant distribution of particles is unstable with respect to formation of rapidity
dependent density fluctuations.  The seeds of these fluctuations arise in the quantum
wavefunctions for the hadrons, and over time become amplified to a magnitude
typical of the Glasma fields.  It is not established whether there is sufficient time in
collisions of nuclei either at RHIC or LHC for this effect to thermalize the produced matter.
This is an area where there is much progress and excitement. 

\section{The Quark Gluon Plasma}

\subsection{Hydrodynamics}

If the matter produced in heavy ion collisions forms a well thermalized Quark Gluon Plasma,
then the evolution of this matter should be well described by perfect fluid hydrdynamics.
There are a variety of approaches.  Some are phenomenological such as the Blast Wave Model.
Some are fundamental, but make assumptions on the initial conditions which are inflexible,
such as the Buda-Lund Hydro Model.  Some such as the SPheRIO model use state of the art
numerical methods and can solve the hydrodynamic equations for arbitrary initial conditions
and equations of state. 

Csorgo presented the state of the art for Buda-Lund Hydro.  This theory is for very specific
initial conditions allows for analytic solutions.  It can be directly compared to the more 
phenomenological results of Blast Wave presented by Kiesal.  Buda-Lund provides a nice
theoretical laboratory where one can study the effects of various equations of state.  It also may
be a fixed point of the more general hydrodynamic solutions at large time.

Grassi presented beautiful results from the SPheRIO model.  She showed that fluctuations
in the initial conditions can affect the extraction of v2, and argued that this may affect
previous extractions. 

Koide and Wolschin presented the state of the art for attempts to include
the affect of viscosity into relativistic hydrodynamics.  This causes problems
with either negative entropy productions or with causality.  It seems that these problems
are controllable.

\subsection{The High Density Quark Gluon Plasma}

\begin{figure}[ht]
    \begin{center}
        \includegraphics[width=0.45\textwidth]{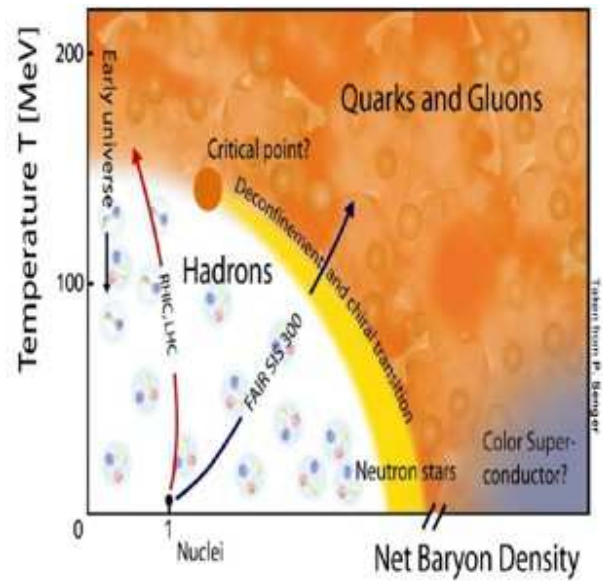}
        \caption{The phase diagram for QCD as a function of baryon density
        and temperature. }
\label{lowdensity}
    \end{center}
\end{figure}

The Quark Gluon Plasma at high baryon number density is the subject of renewed interest.
As shown in Fig. \ref{lowdensity}, there is an expected critical point at some value
of baryon number density and temperature.  Roland showed that experiments
to search for this critical point were feasible at RHIC.  He also presented the
plot of $K^+/\pi^+$ from the SPS heavy ion experiments which shows a sharp peak
at an energy of 5-10 GeV.  This has been argued to be a hint for the critical point, but one should
be cautious as precisely where the peak occurs, one is joining together data from different
experiments.

Lacy argued that either the low $p_T$ $K/\pi$ fluctuation, or a minimum in the viscosity to entropy ratio may
provide a signal for the critical point.  He presented provocative arguments that such a minimum
might be seen.  

\subsection{Jet Energy Loss}

In experiments at RHIC energy, jet energy loss has provided strong indications
what one has produced a strongly interacting Quark Gluon Plasma.  Lajoie presented latest
data from Phenix concerning this energy loss.  One has a good phenomenological
understanding of jet energy loss for light quarks, but a detailed quantitative theory
is difficult.\cite{whitepaper}

One of the outstanding mysteries is the apparent large amount of energy loss of
charmed particles.   This is also related to the large flow of charm.  It appears that
in spite of the large mass of the charmed particle, it slows down in a  media like a light particle.
This is a surprise since in the rest frame of a charmed quark, the typical energy exchange
in a hadronic interaction, should be of the QCD scale, and the fractional energy loss
decreases as the inverse of the charm mass.  Boosting to the fast moving frame of
a charmed quark, the fractional energy loss remains invariant, so the heavy quark
does not slow down much due to a collision.  

Charm quark energy loss was underscored as one of the major problems with jet energy
loss calculations in the talk by Vitev.  He also presented some beautiful calculations which show
the jet suppression factor dependence  on the number of participating nucleons in a collision,
$ln(R_{AA}) \sim -\kappa A^{2/3}$.

Gossiaux emphasized that our lack of understanding of heavy quarks is great.  Predictions
for the cross section of open charm production do not agree with RHIC data.  We
do not have a comprehensive picture of $J/\Psi$ production.  Energy loss and flow data do
not agree with expectations.  

Perhaps some of the problem concerning heavy quark energy loss might be resolved 
if there was a smaller contribution of bottom than expected.  Suade argued that
electro-hadron correlations might be useful here.

Xu argued that the fragmentation of gluon jets and quark jets should be different.  Quark
jets and gluon jets should have different energy loss mechanisms in the Quark Gluon Plasma.
It is therefore mysterious why the fragmentation products in AA collisions appears to be the
same as in pp collisions.

It is difficult to judge how serious the problems are here.  There is still not consensus about how
to compute jet energy loss, and which mechanisms are the dominant one.  Nevertheless,
there are very bright people thinking about these problems, and the field is young.

\subsection{Global Properties of Heavy Ion Collisions}

Fachini gave an excellent overview of the global properties of heavy ion collisions.
She argued that flow is well described by hydrodynamic computations up to transverse
momenta of $1~GeV$.  She argued that perhaps the saturation of hydrodynamic
bounds at RHIC may be accidental and that at higher energies these bounds might be exceeded.
These bounds arise from assuming Glauber type initial conditions, and Color Glass Condensate
initial conditions allow for more flow.  This implies that viscous effects would be non-negligable
at RHIC, if the Color Glass Initial Conditions were used.

Fachini also argued that the flow behaviour at $p_T \ge 1~GeV$ might be explained by
participant scaling.  Here one takes the v2 of an observed particle and divides by the number
of its participants.  Then one takes either the transverse kinetic energy or the transverse momentum
and divides also by the number of participants.  The resulting distribution is universal and
independent of  observed particle up to several GeV.

This can be explained in coalescence models.  Such models however violate energy conservation.
The scaling behaviour is nevertheless remarkably good, better than I would expect from the 
models.  Nevertheless, this basic underlying coalescence mechanisim  is strongly suggested.
 
Fachini also discussed a possible mass shift of the $\rho$ meson.  In peripheral 
$Au+Au$ collisions in Star there is a mass shift but no broadening.  This may be due
to interferences between different channels producing the $\rho$ and is likely a final state effect.
In NA60 central $In+In$ collisions, there is a broadening of the $\rho$ but no mass shift.
Are these measurements in agreement?

Rapp argued that one can understand the $\rho$ broadening in $In+In$ collision at the SPS.
It is due to in media interactions of the $\rho$ meson.

Finally, Fachini analyzed thermal models of particle production.  Thermal models provide a 
remarkably good description of particle ratios.  One can also estimate transverse
flow velocities and temperatures at decoupling by the shapes of $p_T$ distributions.

Wiik argued that there should be fluctuations in the Hadgedorn spectra.  This would put
the SPS data in better agreement with experiment.  Such an adjustment is not required for the 
RHIC data.

\section{Hanberry-Brown-Twiss Interferometry}

By studying the correlations of identical particles, it is possible to experimentally determine
the time and spatial region over which particles stop interacting.  This is the so called
surface of decoupling.  (In fact for an evolving system such as a heavy ion collision,
it is not really a surface, since at each time there is a spread out surface
due to  by fluctuations in the last scattering position, and  the
shape of the surface evolves in time.)

Metzger presented a remarkable analysis of data from the L3 detector at LEP.  By
a very detailed analysis he showed that deviations from a Gaussian parameterization
demonstrated that the the space-time surface of decoupling is consistent with
inside-outside cascade dynamics.  This picture is at the heart of the description
of heavy ion collisions.  A description which incorporates this dynamics is that of Csorgo
and Zimanyi.\cite{csorgo}

Ukleja presented a comparison of of typical size scales associated with LEP and Hera.

Chung presented an analysys which suggest that the non-Gaussian tail of the distribution
measured at RHIC may, as was the case at LEP, provide non-trivial information on
the decoupling surface.  This has the potential to modify conclusions drawn from
a Gaussian analysis.

In my opinion, there is still much to be learned from HBT analysis of heavy ion collisions.
It is clear that Gaussian fits miss much of the physics.  Cramer argued that there may
be coherent scattering combined with absorption at late time in the collision, and that
this can modify the conclusions.   Perhaps some of ideas described by Flowkowski concerning
multiple correlations and jest may be useful in sorting all of this out.

\section{The Highest Energies}

The highest energies collisions observed remain those of cosmic rays.
Licinio argued that one may have access to new physics in such collisions,
but this is difficult since the properties of showers are largely determined by the physics
of the fragmentation region.  So long as there is limiting fragmentation, the shower 
properties are determined.  Nevertheless, Escobar argued that cosmic rays
at the highest energies may allow us to do astronomy.  At the highest energies, 
cosmic rays are no bent much by galactic and extra-glactic magnetic fields.
This allows in principle the identification of the highest energy source of cosmic rays.
Candidates for such sources include active gaactic nuclie (black holes) and neutron stars.

\section{Acknowledgements}

I thank the organizers for inviting me to give this summary.  Also for the wonderful
atmosphere in Paraty.  It was a very good meeting, where the full range
of physics associated with strong interactions at high energies was presented.

This manuscript has been authorized under Contract No. DE-AC02-98CH0886 
with
the U. S. Department of Energy.

\end{document}